# The Dynamic X-ray Sky of the Local Universe
*A White Paper Submitted to the Decadal Survey Committee*

**Contributors:**
Alicia M. Soderberg *(Harvard/CfA)*, Jonathan E. Grindlay *(Harvard/CfA)*, Joshua S. Bloom *(UC Berkeley)*, Suvi Gezari *(Johns Hopkins)*, Anthony L. Piro *(UC Berkeley)*, Tomaso Belloni *(INAF-Brera)*, Jifeng Liu *(Harvard/CfA), Ada Paizis (INAF-Milan),* Edo Berger *(Harvard/CfA)*, Paolo Coppi *(Yale)*, Nobu Kawai *(Tokyo Inst. Tech.),* Neil Gehrels *(GSFC)*, Brian Metzger *(UC Berkeley)*,  Branden Allen *(Harvard/CfA)*, Didier Barret *(CESR-Toulouse)*, Angela Bazzano *(IASF-Rome)*,  Giovanni Bignami *(IASF-Milan)*, Patrizia Caraveo *(IASF-Milan),* Stephane Corbel *(U. Paris Diderot, CEA Saclay)*,  Andrea De Luca *(INAF-Milan)*, Jeremy Drake *(Harvard/CfA)*, Pepi Fabbiano *(SAO)*, Mark Finger *(USRA)*, Marco Feroci *(INAF-Rome),* Dieter Hartmann *(Clemson)*, JaeSub Hong *(Harvard/CfA)*, Garrett Jernigan *(UC Berkeley)*, Philip Kaaret *(U. Iowa)*,  Chryssa Kouveliotou *(MSFC),* Alexander Kutyrev *(GSFC),* Avi Loeb *(Harvard/CfA),* Giovanni Pareschi *(INAF-Brera)*, Gerry Skinner *(GSFC)*, Rosanne Di Stefano *(Harvard/CfA),* Gianpiero Tagliaferri *(INAF-OABrera)*, Pietro Ubertini *(IASF-Rome),* Michiel van der Klis *(U. Amsterdam),* Colleen A. Wilson-Hodge *(MSFC)*

**Science Frontier Panel:** Stars and Stellar Evolution (SSE)
**Facilities:** EXIST, LSST, SKA pathfinders, E-VLA

> **Key Goals:**
> 1. Exploiting a novel technique to pinpoint supernovae at the moment of explosion, revealing the progenitor properties and explosion mechanisms of supernovae and sub-energetic gamma-ray bursts and their utility as beacons for coincident neutrino and gravitational wave searches.
> 2. Closing in on magnetic field generation in stars, from determining the habitability of planets orbiting coronally active stars to revealing the extreme physics underlying magnetar giant flares.
> 3. Pinpointing accreting black holes of stellar and (possibly) intermediate mass based on their transient outbursts, and uncovering quiescent super massive black holes by their tidal disruption of field stars.
> 4. Exploring **new phase space of the Transient Universe** at X-ray wavelengths, in parallel with complementary time domain efforts at longer wavelengths. Enabling in depth studies of known transients, first detections of predicted X-ray phenomena, and exploring the looming class of unknown X-ray transients.

**Abstract**
Over the next decade, we can expect time domain astronomy to flourish at optical and radio wavelengths.  In parallel with these efforts,  a dedicated transient "machine" operating at higher energies (X-ray band through soft gamma-rays) is required to reveal the unique subset of events with variable emission predominantly visible above 100 eV. Here we focus on the transient phase space never yet sampled due to the lack of a sensitive, wide-field and triggering facility dedicated exclusively to catching high energy transients and enabling rapid coordinated multi-wavelength follow-up.  We first describe the advancements in our understanding of known X-ray transients that can only be enabled through such a facility and then focus on the classes of transients theoretically predicted to be out of reach of current detection capabilities.  Finally there is the exciting opportunity of revealing new classes of X-ray transients and unveiling their nature through coordinated follow-up observations at longer wavelengths.

**Motivation**

The local Universe (within a few hundred Mpc) is teeming with transient astronomical phenomena, a distinct subset of which are only identifiable by their high energy emission. Over the past two decades we have made the preliminary steps in mapping the population and demographics of local X-ray transients, enabled by the capabilities of past (CGRO, Einstein, ROSAT, BeppoSAX, HETE-2) and current (RXTE, Integral, Chandra, XMM, Swift) missions. We now know that the bulk of local X-ray transients can be traced to accreting compact objects including stellar mass black holes (BHs), neutron stars (NSs), and white dwarfs (WDs).

More intriguingly, these missions have also revealed a growing sample of rare and exotic X-ray transients in the local Universe, whose nature and driving emission mechanisms remain largely unknown. Chief among these are the newly discovered class of sub-luminous gamma-ray bursts that volumetrically outnumber both the classical long-duration GRBs and short-hard bursts by a factor of ten [1]. However, their weak prompt emission (L ~ $10^{46}$ erg/s; roughly $10^6$ times lower than cosmological GRBs) currently limits their detectability to within 200 Mpc and leaves open the question of their progenitors in the context of other cosmic explosions. Packing a similar X-ray punch are the handful of Giant Flares detected to date from Soft Gamma-ray Repeaters (SGRs), widely attributed to 10-100 TeraGauss NSs ("magnetars") but whose magnetic field production and duty cycle remain poorly understood [2,3].

Also blazing brighter than the Eddington luminosity, $L_{Edd}$ ~ $10^{38}$ erg s$^{-1}$ ( M~M$_\odot$), is the class of Ultra-Luminous X-ray sources (ULXs), found by the dozens in nearby galaxies but strangely absent from our Milky Way. Spawned by fortuitous observations of luminous outbursts ($\Delta L_X$ ~$10^3$ [4]) comes the hotly debated idea that ULXs are powered by a class of intermediate mass BHs (IMBHs) weighing in at 100-1000 times that of the Sun [5].

Progress requires a more sensitive wide-field and triggering X-ray transient mission that will not only sharpen our understanding of known transients in the local Universe, but reveal new classes of outbursting objects some of which are already expected on theoretical grounds. In this spirit, a prime example is the decades-ago predicted X-ray pulse marking the precise moment that a supernova (SN) shock wave breaks out of the progenitor star [6]. This luminous ($L_X$~$10^{44}$ erg/s) beacon acts as a "time stamp" for the explosion, enabling more sensitive searches for coincident gravitational waves [7,8] and neutrinos. While the observational realization of this fleeting signal (sec to min) has long been known to require a dedicated X-ray transient facility, the pivotal first detection was made serendipitously last year with the narrow-field *Swift* X-ray Telescope (XRT) showing properties (luminosity, duration) in line with early expectations [9,10].

With a predicted X-ray luminosity not dissimilar from SN shock breakout is the *prompt* ($\Delta t$~10 sec) tidal compression pulse preceding the disruption of a star around a super massive BH (SMBH; [11]). While X-ray and UV observations have revealed evidence for long-lived fall-back flares from about a dozen tidally disrupted stars to date [12,13,14], the detection of the preceding pulse requires a wide-field and triggering X-ray facility to catch these rare events. Such early warning beacons will ultimately enable mass estimates for a larger and more distant sample of SMBHs, which in turn, hold the greatest promise of extending the existing M-σ relation beyond current limitations by revealing dormant SMBHs in galactic nuclei [15,16].

Finally is the class of unknown transients for which we currently lack both predictions and detections. This class represents a significant area of discovery space that only a wide-field and sensitive X-ray transient "machine" can uniquely explore. In the sections below, we outline the importance of extending our knowledge of known, predicted, and unknown X-ray transients, on par with the on-going ground-based technological efforts to advance our understanding of the dynamic sky at optical (LSST) and radio (SKA pathfinders) wavelengths.

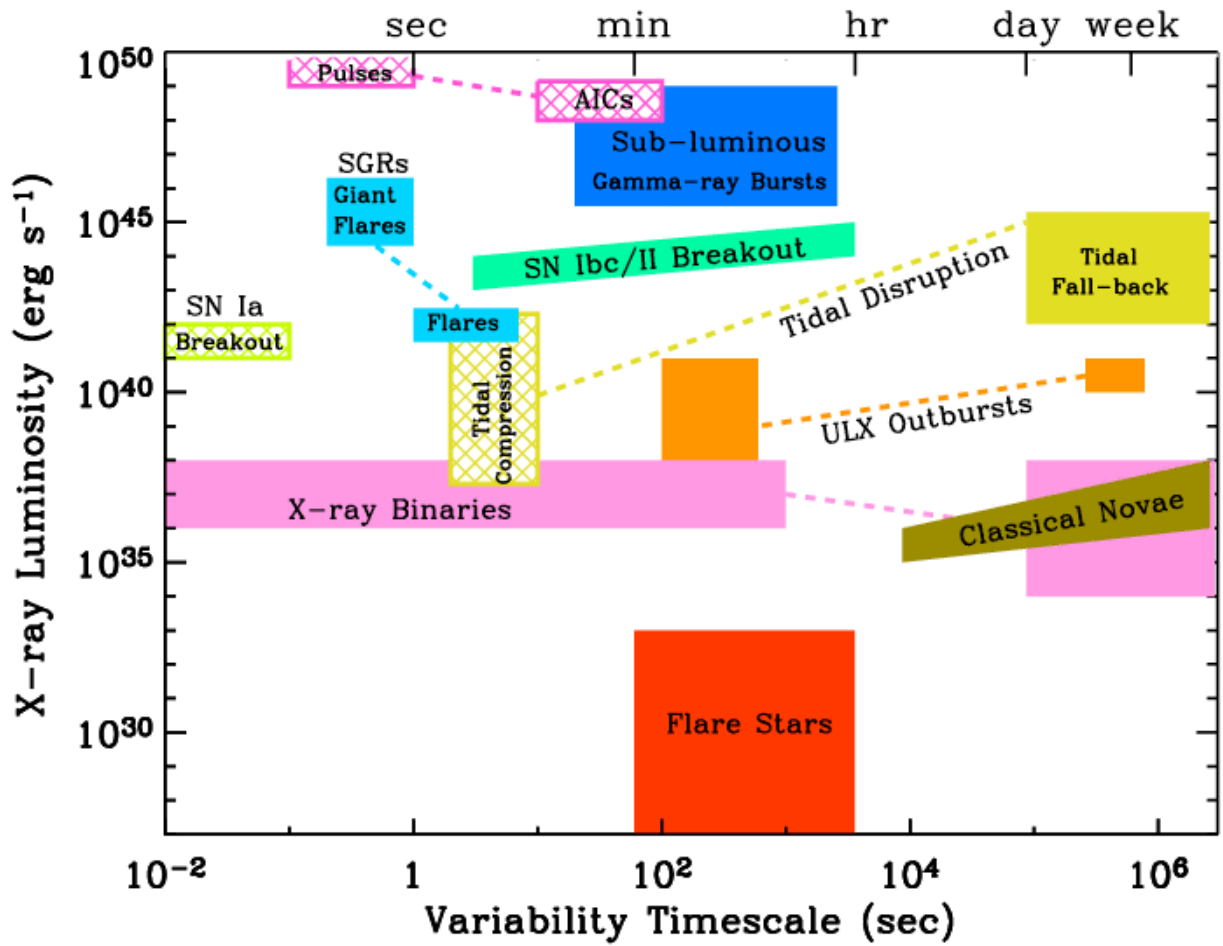

**Figure 1:** *A compilation of high energy transients that are known (solid) and/or predicted (hatched) in the local (d< 200 Mpc) Universe. The variability timescale is the characteristic duration of the transient outburst, plotted here as a function of the peak X-ray luminosity during the outburst. Transients with variability on multiple timescales are linked with dashed lines. Sub-luminous, fast ( < 10 sec), and rare transients have historically been missed due to instrumentation limitations. Characteristics compiled from references [1-6,9-14,16,19-23].*

**A Diversity of X-ray Transients**

The dynamic nature of the local X-ray Universe was established decades ago, revealing diverse classes of outbursting objects (Figures 1 & 2). In particular, those transients with long duty cycles (> months), high peak luminosities ($L_X > 10^{37}$ erg/s) and/or large populations (> 100s per galaxy) were the first to be detected, catalogued and studied. At the same time, some of the most fundamental questions remain unanswered for X-ray transients, largely due to poor localizations and the lack of simultaneous observations at longer wavelengths.

A two-pronged strategy is required to make progress. First, mapping out the luminosity function and duty cycle of rare transients requires larger samples only possible with an X-ray facility with increased sensitivity and a wider field-of-view. Next, a deeper understanding of the emission processes powering the X-ray transients relies on coordinated multi-wavelength observations, which in turn, call for arcsecond localizations, autonomous triggering capability and rapid response on-board follow-up telescopes, and real-time dissemination of trigger alerts to the community. Below we list some of the central questions facing the diverse classes of known local X-ray transients and the critical steps required for progress:

✦*What is the relation between sub-energetic GRBs and classical GRBs? What essential physical process enables a small fraction of SN explosions to produce relativistic material?*
> With the recent discovery of three sub-luminous GRBs at z < 0.1 each associated with a Type Ibc SN, came the realization of a distinct class of explosions that bridge classical GRBs and ordinary SNe [1]. These mildy-relativistic bursts challenge our notion that all GRBs share a common energy reservoir and hint that apparently ordinary SNe may harbor central engines powering weak (and ultimately suffocated) GRB jets. Until the luminosity function of sub-energetic GRBs is well mapped, the nature of the connection between ordinary SN progenitors and those of GRBs remains unclear. A high energy triggering facility with a factor of ~10 increase in sensitivity over Swift/BAT and Integral/IBIS would increase the sample of sub-energetic GRBs significantly. Moreover, on-board follow-up instruments would enable arcsec localization and prompt coverage across the EM afterglow spectrum as well as precise time-stamps required to search for coincident gravitational wave and neutrinos (see [17] for Astro2010 review).

✦ *How are the 10-100 TeraGauss magnetic fields generated in magnetars? What essential process drives the duty cycle of giant flares? Mapping the SGR life cycle from birth to death.*
> Thanks to their powerful X-ray flares and giant flares, a handful of SGRs have been discovered to date, all but one in our galaxy, and with a wide dispersion of luminosities [2,3]. Extending the sample requires a more sensitive X-ray facility to identify magnetars in nearby galaxies using their short-lived (~1 sec) giant flares as beacons. Rapid response follow-up is critical to detect the fading "afterglow" from these explosions, a non-thermal signal that often reveals the underlying slow-spin (~5-10 sec) of the underlying neutron star, and constrains the enormous magnetic field strength. A sensitive and wide-field triggering facility with autonomous on-board prompt follow-up capability is required to increase the sample and distinguish SGR giant flares from short-hard GRBs [17].

✦*Are ULXs powered by accreting stellar mass BHs with "special" emission properties (beaming) or direct evidence for IMBHs? What drives the powerful outbursts seen from some ULXs and what is the nature of the accretion flow?*
> While hundreds of ULXs have already been discovered in the local Universe, coordinated X-ray and high-resolution optical observations are essentially non-existent and yet required for identifications. For the most luminous sources ($L_X > 10^{40}$ erg/s during "high" states) the primary objective is to constrain the mass of the BH. Arcsecond localizations of ULXs are required to identify their faint optical/IR counterparts (B > 22 mag). For transiting systems, optical/IR and X-ray monitoring reveals the period of the orbit and may determine the mass function of the system [e.g., 18]. Finally, high cadence coordinated monitoring can reveal the details of the accretion disk physics (clumpy vs wind-fed) giving rise to the observed short timescale (minutes) modulations and revealing the super outbursts lasting days to weeks.

✦ *Mapping the magnetic activity of stars: what are the implications for planet habitability? What is the luminosity function for stellar X-ray flares and what powers the super outbursts seen from some giant stars?*

Recently, some stars have been seen in super outburst (e.g. RS CVn star II Peg, [19]) tapping into >10% of the total bolometric luminosity and producing strong non-thermal emission from accelerated coronal particles detectable in the radio and hard X-ray bands. To catch these rare fleeting (minutes) signals of which we currently have less than half a dozen examples, a sensitive triggering facility is required. Moreover, coordinated X-ray and radio (E-VLA) follow-up to reveal the details of the magnetic field (scale, strength) by pinning down the broadband non-thermal spectral properties. In turn, the impact of the magnetically powered ionizing flux (including outburst duty cycle) on close orbiting planets can be investigated in the context of habitability.

✦ *What fraction of supersoft X-ray sources are attributed to classical novae and what are the implications for the life cycle and local environments of Type Ia supernova progenitors?*

As the focus of optical astronomy turns towards time-domain studies (e.g., LSST), we can expect the visual-band discovery and monitoring of classical (and recurrent) novae to increase by leaps and bounds. At the same time, recent high energy observations of accreting WD systems reveal a diversity of X-ray counterparts, often in Supersoft Source (SSS) states and some showing extreme variability (e.g., RS Ophiuchi [20]). Coordinated multi-wavelength studies of accreting WD transients are required to shed light on their properties (e.g., mass growth), accretion history, duty cycle and, in turn, the local environments shaped by these eruptions. Such studies have paramount importance in the race to identify Type Ia progenitors.

✦ *What powers the fast flares seen for a small fraction of high-mass X-ray binaries and what does it reveal about the evolution of these systems?*

Thanks to decades of study, the overall diversity of Galactic X-ray binaries and the properties of their accretion-powered high energy emission have been delineated (see [21] for Astro2010 review). Recently, however, rare cases have emerged with observed X-ray characteristics (luminosity, outbursts) distinct from typical Galactic systems. In particular, emitting bright ($L_X \sim 10^{36}$ erg/s), short (min to hr), and sporadic flares are the class of Supergiant Fast X-ray Transients (SFXTs) for which there are currently just a handful of known examples [22]. While the proposed mechanism for the outbursts is largely attributed to variability of the Supergiant wind, the exact method by which the star feeds the compact object (BH, NS, or pulsar), has yet to be revealed and will ultimately shed light on the evolution of high-mass X-ray binaries [23].

**On the Cusp of Discovery: Predicted X-ray Transients**

Limited by the capabilities of current and previous instrumentation, we have been chronically insensitive to those transients, usually detectable in X-rays with low volumetric rates, short-lived and/or sub-luminous emission. On theoretical bases, however, the local

Universe abounds with such transient astronomical phenomena. Below we outline the some of the exciting transients predicted to be hovering just beyond current detectability.

✦ *SN shock breakout outbursts:*
With the serendipitous discovery of shock breakout emission from SN 2008D (d~27 Mpc) caught with the *Swift*/XRT [9] came the long-awaited observational evidence that core-collapse supernovae ($M_{ZAMS} > 8$ $M_\odot$) produce a detectable soft X-ray pulse at the moment the star explodes [6]. Predicted to show a luminosity of $L_X \sim 10^{43}$ to $10^{46}$ erg/s depending on the properties of the progenitor, this fleeting signal (seconds to minutes) should accompany *every* core-collapse SN, at a rate of once per hundred years per galaxy [24]. Interestingly, thermonuclear SNe Ia are also expected to produce a prompt breakout signal although the observable properties ($L_X \sim 10^{42}$ erg/s, $\Delta t \sim 0.01$ sec) are significantly more difficult to detect owing to the WD progenitor's small radius and low mass [25].

The sensitivity of current narrow-field X-ray missions restricts the detectability of SN shock breakout bursts to within a few dozen Mpc, framing the observational challenge of knowing "when and where to look". With the discovery of SN 2008D, we now know that a wide-field sensitive and triggering satellite could uncover dozens of SNe in the local Universe each year in the act of exploding. The long history of relying on strong optical emission to discover SN will give way to this efficient prompt detection method; soon every SN within the local Universe could enjoy multi-wavelength follow-up with breadth rivaling SN 2008D and even SN 1987A.

Of special note, the use of SN breakout X-rays can provide a precise time-stamp for the gravitational collapse of the star (accurate to within minutes of breakout), enabling searches for coincident gravitational waves and neutrinos that, in turn, encode the explosion physics [7,8]. Both signals are expected to be strongest in the case of asymmetric core-collapse SNe, although in-spiraling double-degenerate WDs (the possible progenitors of SNe Ia) should also produce a strong gravitational chirp.

✦ *Tidal Compression Flares:*
While X-ray and UV observations have revealed evidence for long-lived (weeks to months) fall-back flares from about a dozen tidally disrupted stars to date, a sensitive X-ray survey with fast triggering capability could enable detection of the predicted prompt pulse of emission from stellar compression. With a luminosity of $L_X \sim 10^{39}$ erg/s, and duration of $\Delta t \sim 10$ sec (for $M_{BH} \sim 10^6$ $M_\odot$) this never-yet detected early warning beacon could be seen with projected wide-field X-ray telescopes out to 10 Mpc. Meanwhile, triggered detection of the more luminous fall-back phase X-ray emission can enable rapid follow-up at IR/optical/UV wavelengths to trace the evolution, and in turn, reveal the mass (and perhaps the spin) of the SMBH. Searches for coincident gravitational wave (GW) signals can also be enabled through accurate event timing, particularly for disrupted sub-giant stars whose WD cores will inspiral, producing a strong chirp. In this context, tidal disruption signals hold promise for revealing the demographics of dormant SMBHs [15] and extending the M-σ relation to larger distances.

✦ *Accretion Induced Collapse Outbursts:*
> It is largely accepted that some massive accreting WDs (those with O/Ne/Mg cores) do not give rise to thermonuclear Type Ia supernovae, instead collapsing inward upon their dense cores [26]. This accretion induced collapse (AIC) leads to the formation of a NS, and in some cases, a magnetar [27]. It is estimated that about $10^{49}$ to $10^{50}$ erg of energy is liberated in this process, and may power a short ($\Delta t < 1$ sec) pulse of high energy emission and thus be responsible for at least some short-hard GRBs. Moreover, it is predicted that the signal may be followed by extended emission (up to 2 min) powered by rotational energy of the newly-born magnetar. Observations of these processes lead to direct constraints on the WD density profile, explosion process, and shed light on the life cycle of NSs and magnetars. The volumetric rate of AICs is predicted to be low, roughly 100 times less common than SNe Ia, or about once per $10^{4-5}$ years per galaxy and comparable to the (poorly determined) rate of SHBs. However, thanks to their theoretically argued strong X-ray signals, AICs could be detected with a few hundred Mpc with a wide-field, sensitive and triggering mission.

**The Class of Unknown X-ray Transients**

Perhaps the most exciting aspect of a X-ray transient "machine" is the possibility of revealing previously unknown transients for which we have no previous theoretical expectations. This unique discovery space is well matched by that at optical/IR (PanSTARRS and LSST) and radio (E-VLA, SKA pathfinders) wavelengths, accessed through synergistic coverage -- both for discovery and rapid response follow-up -- to lift the veil on the dynamic local Universe. Non-thermal and thermal emitters will be uncovered by the combination of these surveys and the cross-talk between them and other telescopes, specifically GW and neutrino, will reveal diagnostics on the EM and non-EM signals from transient phenomena.

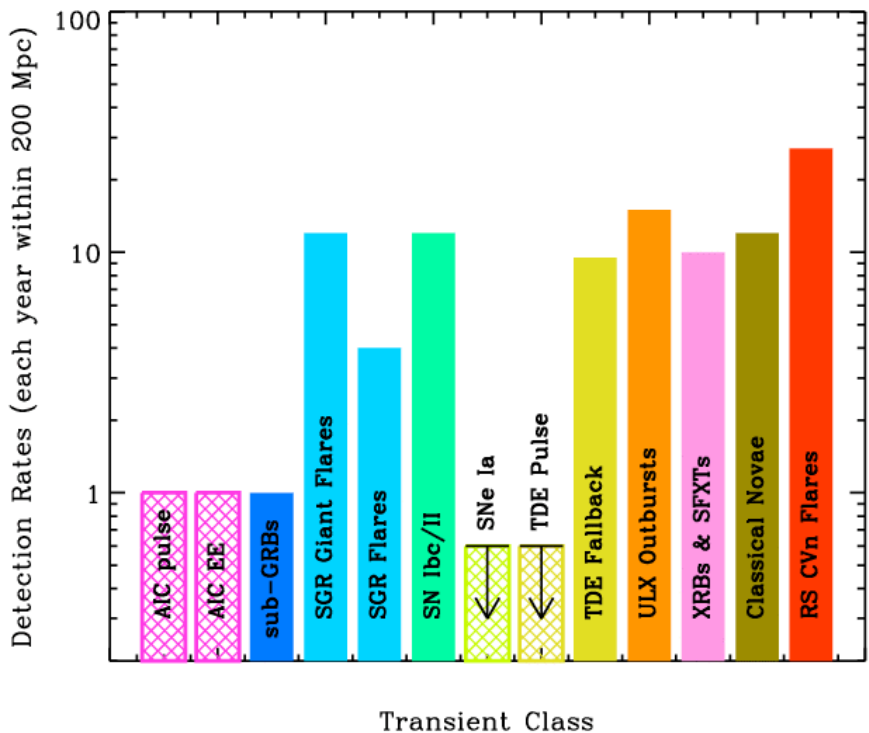

**Figure 2:** *The estimated local detection rate (per year, within 200 Mpc) of known (solid) and predicted (hatched) X-ray transients discussed in the text and ordered according to luminosity shown in FIgure 1. A sensitive, wide-field (2 sr), triggering mission with sensitivity a factor of ~10 greater than that of Swift/BAT and Integral/IBIS and comparable to that of the high energy telescope planned for EXIST will facilitate progress by increasing the sample size and enabling for prompt multi-wavelength follow-up from on-board and ground-based telescopes. Rates compiled from [1-6,9-11,16-19,21,24-29].*

**Relevant Facilities**

The Energetic X-ray Imaging Survey Telescope (*EXIST*)[1] is an AMCS X-ray imaging all-sky deep survey mission. *EXIST* is based on proven technology and would image and temporally resolve the entire sky every two 95-minute orbits, detecting extremely faint high energy sources in the range (5-600 keV). Most important, *EXIST* has an unparalleled combination of localization (20 arcsec, 5σ), triggering capability, wide-field sensitivity ($F_X \sim 2 \times 10^{-10}$ $(t/10\ sec)^{-1/2}$ erg cm$^{-2}$ s$^{-1}$; 5σ), the critical requirements for a next generation X-ray transient machine (Figure 2). Finally, *EXIST* has the ability to autonomously slew (< 100 sec) to newly identified transients for prompt coordinated follow-up with two on-board narrow-field instruments including sensitive soft X-ray (0.1-10 keV) and UV/optical/NIR (0.3-2.0 μm) telescopes in addition to disseminating transient alerts to other facilities (e.g., E-VLA, LSST) in real time.

---

[1] http://exist.gsfc.nasa.gov/